\documentclass[12pt]{article}

\usepackage[english]{babel}
\usepackage{amssymb,latexsym,amsmath,mathrsfs}

\begin{document}

\title
{Nonassociative gauge fields}
\author
{E.K. Loginov\footnote{{\it E-mail address:} ek.loginov@mail.ru}\\
\it Ivanovo State University, Ivanovo, 153025, Russia}
\date{}
\maketitle
\begin{abstract}
In this paper, we study consequences of the assumption that the gauge group $SU(2)$ of the standard model is a nonassociative image of $Spin(3)$. Such an approach allows us to take a different look at the Higgs mechanism and obtain the value of the Weinberg angle in very good agreement with the experiment.
\end{abstract}

\section{Introduction}

There are a huge number of grand unification theories that predict the values of various arbitrary parameters of the standard model (see, e.g., the review~\cite{tana18}). A characteristic feature of all these theories is the presence of new, not yet discovered gauge bosons associated with generators of the unification gauge groups. It can be assumed that these particles are not discovered due to the insufficient energy of accelerators, but this assumption encounters certain difficulties. In the standard model, the scale of electroweak symmetry breaking acquires quadratically diverging radiation corrections, the ultraviolet cutoff of which is of the order of the scale of the grand unification theory. In view of the enormous difference between these two scales, it becomes necessary to fine-tuning of parameters of the theory, which looks unnatural. This is usually referred to as the gauge-hierarchy problem~\cite{gild76}.
\par
The issue is whether one can find a model that avoids this unnatural feature. The best-known realization of this approach is based on supersymmetry that cancels quadratic divergences in expressions for the running parameters of the standard model~\cite{marc90,barg93,lang93,schm05}. More radical are models in which the ultraviolet cutoff is of the order of the electroweak theory scale. This is usually achieved by adding a large number of new fields and giving up the perturbation~\cite{ghil97}. However, in both cases, the problem associated with the existence of new particles remains unresolved.
\par
This problem can be drastically solved by constructing a model in which the unification scale is of the order of electroweak, and new particles are absent. This idea was realized in the $SU(3)\times SU(3)$ and $G_2$-models of gauge-Higgs unification~\cite{anto01,csak03}. However, the Weinberg angle predicted in these models is very far from the experimental one. At the same time, a certain modification of the idea of gauge-Higgs unification allows us to solve the problem of gauge hierarchies and obtain predictions that consistent with the experiment.

\section{Preliminaries}

We recall that the algebra of octonions $\mathbb O$ is a real linear algebra with the canonical basis $1,e_{1},\dots,e_{7}$ such that
\begin{equation}\label{01}
e_{i}e_{j}=-\delta_{ij}+c_{ijk}e_{k},
\end{equation}
where the structure constants $c_{ijk}$ are completely antisymmetric and nonzero as $c_{123}=c_{145}=c_{176}=c_{246}=c_{257}=c_{347}=c_{365}=1$. However, in our case it is more convenient to use an alternative definition of the octonion algebra. Let $\mathbb{H}$ be the quaternion algebra with the standard basis $1,e_{1},e_{2},e_{3}$ such that $e_{i}e_{j}=-\delta_{ij}+\epsilon_{ijk}e_{k}$. Then the vector space $\mathbb{H}\oplus\mathbb{H}e$ is called the octonion algebra if we define  on it the multiplication
\begin{equation}\label{02}
(a_1+a_2e)(b_1+b_2e)=(a_1b_1-\bar{b}_2a_2)+(a_2\bar{b}_1+b_2a_1)e,
\end{equation}
where $\bar{b}_i$ is the conjugate to $b_i$ quaternion. In order to reconcile the definitions (\ref{01}) and (\ref{02}), it is enough to put $e_{n+4}=e_{n}e$.
\par
Following~\cite{scha52,mccr66}, we define the notion of representation in the octonion algebra. Let $L_a$ and $R_a$ be the operators of left and right multiplication by the element $a\in\mathbb{O}$. Then a pair of linear mappings $(L,R):\mathbb{O}\to\text{End}\,\mathbb{O}$ of the octonion algebra into the endomorphism algebra of the linear space $\mathbb{O}$ is called the regular representation of the algebra $\mathbb{O}$. The restriction of the regular representation of the algebra $\mathbb{O}=\mathbb{H}\oplus\mathbb{H}e$ to the subalgebra $\mathbb{H}$ gives a regular representation of the quaternion algebra. It is defined by the following system of equalities
\begin{equation}\label{03}
L_{ab}=L_aL_b,\quad L_aR_b=R_bL_a,\quad R_{ab}=R_bR_a.
\end{equation}
A representation satisfying the system of equalities (\ref{03}) is called associative. In particular, if all the operators $R_{a}$ in ((\ref{03}) act trivially, then we have the usual (left) representation of associative algebra.
\par
An alternative representation example can be constructed if we consider the subspace $\mathbb{H}e$ in $\mathbb{O}$ and define the action of $\mathbb{H}$ on $\mathbb{H}e$ as a restriction of the regular representation of the algebra $\mathbb{O}$. In this case, it follows from the law of multiplication (\ref{02}) that the action of $\mathbb{H}$ on $\mathbb{H}e$ is determined by the equalities
\begin{equation}\label{04}
L_{ab}=L_bL_a,\quad R_a=L_{\bar{a}}.
\end{equation}
Since this representation does not satisfy the system of identities (\ref{03}), it is called non-associative.
\par
Now let $G$ be the set of all elements of norm 1 in the quaternion algebra. Since it coincides with the Clifford algebra $Cl_{0,2}(\mathbb{R})$, we may assume that $G=Spin(3)$. Then the equalities (\ref{03}) and (\ref{04}) induce the associative $G\to\text{Aut}\,\mathbb{H}$ and the non-associative $G\to\text{Aut}\,\mathbb{H}e$ representations of $G$ respectively. In turn, the representations of the group $G$ induce representations of its tangent Lie algebra $A_G$. It is easy to obtain these representations if we consider $A_G$ as a subalgebra of the commutator algebra of $\mathbb{H}$. Let the representation of $G$ be associative. Suppose $T_a=L_a-R_a$ and consider the map $T:A_G\to\text{End}\,\mathbb{H}$. Then it follows from (\ref{03}) that
\begin{equation}\label{13}
T_{[a,b]}=[T_a,T_b].
\end{equation}
This homomorphism defines an ordinary Lie representation of $A_G$. Now let the representation of $G$ be nonassociative. Suppose again $T_a=L_a-R_a$ and consider the map $T:A_G\to\text{End}\,\mathbb{H}e$. Then it follows from (\ref{04}) that $R_a=-L_a$ and therefore
\begin{equation}\label{11}
T_{[a,b]}=\frac{1}{2}[T_b,T_a].
\end{equation}
Since this representation does not satisfy the system of identities (\ref{13}), it is called the non-Lie (Mal'tsev) representation.

\section{Higgs mechanism}

To construct a gauge-invariant Lagrangian, it is usually required that the covariant derivative $D_{\mu}\phi$ have the same transformation properties as the field $\phi$. This condition determines the well-known law of transformation of gauge fields
\begin{equation}\label{21}
A'_{\mu}=uA_{\mu}u^{-1}+u\partial_{\mu}u^{-1},
\end{equation}
whose form does not depend on the field $\phi$ in the representation space of the gauge group. However, this is true only when considering associative representations for which $(A_{\mu}u)\phi=A_{\mu}(u\phi)$. Otherwise, the transformation law of the gauge fields takes a different form.
\par
Let $G$ be the set of all elements of norm 1 in the quaternion algebra $\mathbb{H}$. We consider the action of $G$ on $\mathbb{H}e$ defined by the equalities (\ref{04}) and we will find the matrices $L_{e_k}$ representing the generators $e_k$ of $G$. To do this, we put $\phi_1=\theta_0+\theta_3e_3$ and $\phi_2=\theta_2-\theta_1e_3$ and consider the action of $e_k$ on $(\phi_1+\phi_2e_2)e$ on the left. Using (\ref{02}) it is easy to show that $\phi_1$ and $\phi_2$ are transformed exactly like the complex doublet $\phi=(\phi_2,\phi_1)^{t}$ under the action of $2\times2$ matrices
\begin{equation}\label{16}
L_{e_k}=(-1)^ki\sigma_k,
\end{equation}
where there is no summation over $k$ and $e_3$ plays the role of the imaginary unit. Thus, we constructed the anti-isomorphism $L:G\to SU(2)_L$ under which the action of $G$ on $\mathbb{H}e$ induces the action of $SU(2)_L$ on $\phi$.
\par
On the other hand, the doublet $\phi=\phi(x)$ can be identified with the Higgs doublet. Indeed, consider the gauge-invariant Lagrangian
\begin{equation}
\mathscr{L}\label{18}
=-\frac{1}{4}F^k_{\mu\nu}F^{k\mu\nu}+(D_{\mu}\phi)^{\dag}D^{\mu}\phi+m^2\phi^{\dag}\phi-\frac{m^2}{v^2}(\phi^{\dag}\phi)^2,
\end{equation}
where the covariant derivative $D_{\mu}\phi=\partial_{\mu}\phi+L_{A_{\mu}}\phi$ and the field $A_{\mu}=gA^k_{\mu}e_k$. We introduce the polar coordinates for scalar fields and represent the doublet in the form
\begin{equation}
\phi=L^{-1}_u\phi',\quad\phi'=\begin{pmatrix}0\\\frac{v+\lambda}{\sqrt{2}}\end{pmatrix},
\end{equation}
where $u(x)\in G$ the real field $\lambda(x)$ have zero vacuum expectation value. Now suppose that the transformation $\phi\to\phi'$ is infinitesimal, i.e. let $u\approx1+\theta$ and $\theta=g\theta^ke_k$. Then it follows from (\ref{16}) that the doublet
\begin{equation}\label{15}
\phi=\frac{1}{\sqrt{2}}\begin{pmatrix}0\\v+\lambda\end{pmatrix}
-\frac{gv}{\sqrt{2}}\begin{pmatrix}\theta^2-i\theta^1\\i\theta^3\end{pmatrix}+\dots,
\end{equation}
where nonlinear field terms are replaced by dots. Substituting (\ref{15}) in (\ref{18}) and again using (\ref{16}), we obtain
\begin{equation}\label{19}
\mathscr{L}=-\frac{1}{4}\left(\partial_{\mu}\tilde{A}^k_{\nu}-\partial_{\nu}\tilde{A}^k_{\mu}\right)^2
+\frac{g^2m^2}{2}\tilde{A}^k_{\mu}\tilde{A}^{k\mu}+\frac{1}{2}(\partial_{\mu}\lambda)^2-m^2\lambda^2+\mathscr{L}_{I},
\end{equation}
where $\mathscr{L}_{I}$ is the interaction Lagrangian and $\tilde{A}^k_{\mu}=A'^k_{\mu}-\partial_{\mu}\theta^k$. It is clear that the free Lagrangian in (\ref{19}) describes massive vector bosons with masses $gv$ and a scalar meson with mass $2m$. Obviously, this is the usual Higgs mechanism to theories with the $SU(2)$ gauge symmetry and a complex doublet.
\par
Further, the transformation $\phi\to\phi'$ induces the infinitesimal transformation of the vector fields
\begin{equation}\label{20}
L_{A'_{\mu}}=L_{A_{\mu}}+[L_{\theta},L_{A_{\mu}}]-L_{\partial_{\mu}\theta}.
\end{equation}
Using the identity (\ref{11}) and passing from the matrix representation to the fields themselves, we obtain
\begin{equation}\label{17}
A'_{\mu}=A_{\mu}-[\theta,A_{\mu}]-\partial_{\mu}\theta.
\end{equation}
The infinitesimal transformation laws (\ref{20}) and (\ref{17}) do not coincide, they differ in signs before the second term on the right. Therefore, finite transformations induced by them will also differ. If the formula (\ref{20}) induces a finite gauge transformation, then (\ref{17}) results in a transformation which differs from (\ref{21}). To get it, it is enough to rewrite the matrix equality (\ref{21}) in the form
\begin{equation}\label{23}
A'_{\mu}\psi=u(A_{\mu}(u^{-1}\psi))+u((\partial_{\mu}u^{-1})\psi),
\end{equation}
where $u\in G$ and $\psi\in\mathbb{H}e$, and then to use the identity (\ref{02}) with $a_2=b_1=0$. As a result, we get the following transformation:
\begin{equation}\label{22}
A'_{\mu}=u^{-1}A_{\mu}u-u^{-1}\partial_{\mu}u.
\end{equation}
\par
Note that the transformations (\ref{17}) and (\ref{22}) of the vector fields $A_{\mu}$ are not gauge. (Otherwise, the equation $[\xi+\theta,A_{\mu}]=\partial_{\mu}(\xi-\theta)$ must have a solution $\xi=\xi(\theta)$ for all $A_{\mu}\in\mathbb{H}\setminus\mathbb{R}$, which is impossible due to the noncommutativity of the quaternion algebra.) At the same time, their connection with the gauge transformations is quite transparent. If we substitute $u=1+\theta$ in (\ref{23}) and use the complete antisymmetry of the associator $(\theta,A_{\mu},\psi)$, we get
\begin{equation}\label{24}
A'_{\mu}=\{A_{\mu}+[\theta,A_{\mu}]-\partial_{\mu}\theta\}-2(\theta,A_{\mu},\psi)\psi^{-1}.
\end{equation}
It is easy to show that this equality coincides with (\ref{17}). The expression in curly braces is the standard gauge transformation of the field $A_{\mu}$. The last term appears as a consequence of the nonassociativity of the representation $ Spin(3)\to SU(2)$. For this reason, we will call the field $A_{\mu}$ a nonassociative gauge field.
\par
On the other hand, despite the fact that the transformation laws (\ref{20}) and (\ref{17}) are different, the transformations of the ``number fields'' (i.e. coefficients in front of algebra generators) that are obtained from them are identical. Therefore the Lagrangian $(\ref{19})$ looks exactly the same as the regular one for a standard gauge field. Its symmetry is ordinary $SU(2)$ that acts in a usual way. The only difference between this Lagrangian from the regular one is in the numerical value of the coupling constant witch is due to the different normalization of generators of $Spin(3)$ and $SU(2)$. Hence, instead of the $SU(2)$ gauge theory with the complex doublet, we can consider the equivalent to it the $Spin(3)$ field theory with nonassociative gauge fields and a scalar field taking values in $\mathbb{H}e$. Obvious that such transition keeps the values of the numerical fields unchanged, though one leads to a change in the normalization of generators of the groups and the coupling constants.

\section{Coupling constants}

Let again $G$ be the group of elements of norm 1 in the quaternion algebra and $A_G$ be its Lie algebra. Further, suppose    $A_{\mu}=g_2A^{k}_{\mu}e_k$ is a vector field that takes values in $A_G$ and $\tilde{A}_{\mu}=(-1)^kig_2A^{k}_{\mu}\sigma_k$ is its anti-isomorphic image in $su(2)_L$. In the algebra $ A_G $, the Euclidean scalar product $\frac{1}{2}(\bar{e}_ie_j+\bar{e}_je_i)=\delta_{ij}$ is  defined. With the mapping $e_k\to(-1)^ki\sigma_k$,  it goes into the Killing form $\text{Tr}(\sigma_i\sigma_j)=2\delta_{ij}$. Comparing these two scalar products, we see that the normalizations of the generators of $G$ and $SU(2)_L$ are different in $\sqrt{2}$ times.
\par
Consider the vector field $B_{\mu}=g_1A^{0}_{\mu}e_0$, where $e_0$ is a generator of $U(1)$. In order to relate the constants $g_1$ and $g_2$ we will look for an embedding of the Lie algebra $su(2)_L\oplus u(1)$ in a simple compact Lie algebra. To this end, on the (real) vector space $V=\mathbb{C}\otimes\mathbb{O}$ we define the multiplication
\begin{equation}\label{05}
(a+e_0b)(c+e_0d)=(ac-b^{\ast}d)+e_0(bc+a^{\ast}d),
\end{equation}
where $(a_1+a_2e)^{\ast}=a_1-a_2e$. This multiplication must satisfy the following two conditions. Firstly, the generators of $G$ commute with $e_0$ and secondly $U(1)$ acts on the doublet $\phi$ as in the standard model. The first statement is obvious. In order for the second to be fulfilled, we well require that $ae_0=ae_3$ for $a=(\phi_1+\phi_2e_2)e$. In this case, $\phi_1$ and $\phi_2$ are transformed under the action of $e_0$ in the same way as the complex doublet $\phi=(\phi_2,\phi_1)^t$ under the action of the scalar $2\times2$ matrix $R_{e_0}=-i\sigma_0$.
\par
Now we consider the algebra of multiplications $M(V)$, which is generated by all operators of left and right multiplication by the element $a+e_0b$.
From the law of multiplication (\ref {05}) it is easy to find the general form of such operators
\begin{equation}\label{06}
L_{a+e_0b}=\begin{pmatrix}L_a&-L_{b^{\ast}}\\L_b&L_{a^{\ast}}\end{pmatrix},\quad
R_{a+e_0b}=\begin{pmatrix}R_a&-R_bI\\R_bI&R_a\end{pmatrix},
\end{equation}
where the operator $I$ acts on the elements of $\mathbb{O}$ according to the formula $Ia=a^{\ast}$. Let us prove the following proposition.
\medskip\par
\textsc{Proposition 1.} \textit{The algebra $M(V)$ is isomorphic to the algebra $M_{16}(\mathbb{R})$ of all real matrices of size $16\times16$}.
\medskip\par
Indeed, using (\ref{02}) it is easy to show that $(ab)^{\ast}=a^{\ast}b^{\ast}$. Therefore, the map $a\to a^{\ast}$ is an automorphism. But it is well known (see~\cite{mccr66}) that any automorphism of the octonion algebra is internal. Therefore, the operator $I$ is generated by the operators of left and right multiplication on $\mathbb{O}$. On the other hand, $L_eR_e(a+e_0b)=a-e_0b$. Therefore, $M(V)$ contains the matrix $\text{diag}(1,\dots,1,-1,\dots,-1)$. This and the formulas (\ref{06}) imply that the algebra $M(V)$ contains elements
\begin{equation}\label{07}
\begin{pmatrix}0&-1\\1&0\end{pmatrix},\quad \begin{pmatrix}R_{e_k}&0\\0&-R_{e_k}\end{pmatrix}.
\end{equation}
Obviously, these matrices are mutually independent and anti-commutative. It follows that the algebra $M(V)$ is a homomorphic image of the Clifford algebra $Cl_{0,8}(\mathbb{R})$. Since the latter is isomorphic to the simple algebra $M_{16}(\mathbb{R})$, this homomorphism is an isomorphism.
\medskip\par
We denote by $V^{(-)}$ the commutator algebra of the algebra $V=\mathbb{C}\otimes\mathbb{O}$. Then it follows from the condition $ae_0=ae_3$ defined above that $V^{(-)}$ contains the half-direct sum $\tilde{M}=u(1)\dotplus M$ of the Lie algebra $u(1)$ and the non-Lie Malcev algebra $M=\{a\in\mathbb{O}\mid a+\bar{a}=0\}$. Suppose $L(\tilde{M})$ is a subalgebra of the Lie algebra $(\text{End}\,V)^{(-)}$, generated by the operators $L_{\tilde{a}}$, where $\tilde{a}\in\tilde{M}$. Then the following proposition is true.
\medskip\par
\textsc{Proposition 2.} \textit{The Lie algebra $L(\tilde{M})$ coincides with $so(9)$}.
\medskip\par
Let us proceed to the proof of this statement. In the notation (\ref{06}), the algebra $L(\tilde{M})$ is generated by the set $L_{\tilde{M}}=\{L_{e_k+0e_0},L_{0e_k+e_0}\}$, where $k\ne0$. Since the blocks of $L_{e_k+0e_0}$ are generators of $SO(8)$, the matrices $L_{e_k+0e_0}$ are antisymmetric and traceless. Therefore $L(\tilde{M})$ contains $so(8)$. Since $L_{0e_k+e_0}$ is also antisymmetric and traceless, $so(8)$ is a proper subalgebra of $L(\tilde{M})$. On the other hand, it follows from Proposition 1 that $L(\tilde{M})\subseteq so(9)$. Using the fact that the maximal Lie subalgebra of $so(9)$ is conjugate either to the subalgebra $so(r)\oplus so(9-r)$ or to the subalgebra $so(3)\otimes so(3)$, we prove the proposition.
\medskip\par
Now let $g$ and $g'$ be the electroweak gauge coupling constants. It follows from Proposition 2 that the algebra $\tilde{M}$ has a spinor 16-dimensional non-Lie representation $\tilde{M}\to so(9)$, $\tilde{a}\to L_{\tilde{a}}$. We place one family of quarks and leptons in this representation. Since for the simple compact non-Abelian group normalization of generators is fixed by the nonlinear commutation relations of its Lie algebra,  we have $g=g_2=g_1$ and $g_1L_{e_0}=g'Y$, where $Y$ is the weak hypercharge of the multiplet. On the other hand, the definition of the coupling constant $g_2$ depends on the normalization of the generators of $G$. Since the norm of generators of $G$ is $\sqrt{2}$ times less than the norm of generators of $SU(2)_L$, the relation between the coupling constants $g'$ and $g_1$ is  non-canonical and should have the following form
\begin{equation}\label{08}
g_1=\sqrt{\frac{10}{3}}g'
\end{equation}
in the unification scale $M_0$.
\par
In conclusion, a few words should be said about the role of the Lie algebra $so(9)$ in our construction. The need for its use was due to the following two reasons. First, it was necessary to construct a 16-dimensional spinor representation of $\tilde{M}$, and secondly, it was necessary to show that the generators of this algebra can be normalized in the same way. Both of these requirements are automatically fulfilled as soon as the representation $\tilde{M}\to so(9)$, $e_{k}\to L_{e_{k}}$ is built. The latter allows rewriting the vector fields $A_{\mu}$, $B_{\mu}$, $\phi$ in matrix form and associating them with bosons of the Standard Model. It should be noted that this representation is non-exact. In fact, we construct a nontrivial representation of the higher algebra from the generators of its subalgebra. In this case, only $L_{e_{k}}$ are associated with free particles. Other generators of $so(9)$ have the form $[L_{e_{i}},L_{e_{j}}]$ and therefore are associated with nonlinear combinations of fields $A_{\mu}$, $B_{\mu}$ and $\phi$, which cannot be free particles.
\par
Moreover, it is not at all necessary to introduce the algebra $so(9)$ into our construction. The 16-dimensional spinor representation of $\tilde{M}$ can be obtained by constructing the representation $\tilde{M}\to Cl_{0,8}^{(-)}$. The proof of the possibility of the same normalization of generators of $\tilde{M}$ can also be carried out without invoking the properties of $so(9)$. It is enough to define on $\tilde{M}$ the scalar product $(e_i,e_j)=\text{tr}(R_{e_{i}}R_{e_{i}})$. Then from the properties of the octonion algebra and the equality $ae_0=ae_3$ it follows that $(e_i,e_j)=\delta_{ij}$. Thus, the use of the Lie algebra $so(9)$ in our construction is of a purely technical nature.

\section{Unification scale}

The condition (\ref{08}) are valid for the energy scale $\mu\geq M_{0}$. Now we study the regime $\mu<M_0$. The evolution of the electroweak gauge coupling constant controlled by the one-loop renormalization group equation
\begin{equation}\label{09}
\frac{d\alpha_{n}^{-1}(\mu)}{d\ln\mu}=\frac{b_{n}}{6\pi},
\end{equation}
where
\begin{align}
b_1&=-2N_f-\frac{3}{10},\\
b_2&=22-2N_f-\frac{1}{2},
\end{align}
$N_{f}$ is the number of quark flavors, and $\alpha_n^2=g_n^2/4\pi$. We have ignored the contribution coming from higher-order effects since they does not affect the final result. Expressing the low-energy couplings in terms of more familiar parameters, we can represent the solutions of Eq. (\ref{09}) as
\begin{align}
\alpha^{-1}(\mu)\sin^2\theta_{\mu}&=\alpha^{-1}(M_0)-\frac{b_2}{6\pi}\ln\frac{M_0}{\mu},\\
\frac{3}{10}\alpha^{-1}(\mu)\cos^2\theta_{\mu}&=\alpha^{-1}(M_0)-\frac{b_1}{6\pi}\ln\frac{M_0}{\mu}.
\end{align}
Combining these equations and supposing $\mu=M_W$, we obtain
\begin{equation}\label{10}
\sin^2\theta_{W}=\frac{3}{13}\left(1-\frac{109\alpha_W}{9\pi}\ln\frac{M_{0}}{M_W}\right).
\end{equation}
Using the tree-level mass relation
\begin{equation}\label{12}
\sin^2\theta_{W}=1-\frac{M^2_{W}}{M^2_{Z}}
\end{equation}
and the experimental data (see~\cite{tana18})
\begin{align}
M_{Z}&=91.1876\pm 0.0021\,\,\text{GeV},\\
M_{W}&=80.379\pm 0.012\,\,\text{GeV},\\
\alpha^{-1}_{W}&=128.029\pm 0.010\,\,\text{GeV},
\end{align}
we get
\begin{equation}
M_{0}=245.4\pm 8.3\,\,\text{GeV}.
\end{equation}
Thus for the considered model, it can be argued that the unification scale $M_{0}$ is coincided with the vacuum expectation value
\begin{equation}
v=246.2204\pm0.0005\quad\mathrm{GeV}.
\end{equation}
Conversely, substituting $\alpha_{W}$ and $M_{0}=v$ into (\ref{10}) and $M_{Z}$ into (\ref{12}), we obtain
\begin{equation}\label{14}
M_{W}=80.3802\pm0.0018\quad\mathrm{GeV}.
\end{equation}
This excellent agreement with the experimental results. Substituting (\ref{14}) into (\ref{10}), we find the value of the Weinberg angle which exactly matches that given by the tree-level formula (\ref{12}).

\section{Conclusion}

Despite the fact that the idea of gauge-Higgs unification was constantly present in the present work, we cannot say that its result was the construction of a new model. In fact, we were dealing with the standard model and studied only consequences that a small modification of the latter leads to.
\par
One of these changes is associated with the replacement of the gauge group $SU(2)$ by $Spin(3)$. Since these groups are isomorphic, such replacement keeps the values of the numerical fields and coupling constants unchanged, however, it leads to a change in the normalization of generators of these groups. The latter, in turn, shifts the scale of the grand unification to the scale of electroweak symmetry breaking.
\par
The second change of the standard model is more significant. We considered the group $SU(2)$ as a non-associative image of $Spin(3)$. This made it possible to build the unification group without free parameters with which new gauge bosons could be associated. The scale of the unification unexpectedly turned out to be equal to the electroweak. This made it possible to obtain the Weinberg angle in very good agreement with the experiment.

\end{document}